\documentstyle[preprint,aps,epsf,floats]{revtex}

\begin{document}
\setlength\baselineskip{20pt}

\preprint{\tighten \vbox{\hbox{CALT-68-2301}
        \hbox{hep-th/y0010204} 
}}

\title{ Generalized $\star$-Products, Wilson Lines and the Solution
of the Seiberg-Witten Equations \\  }

\author{Thomas Mehen\footnote{mehen@mps.ohio-state.edu}}
\address{\tighten {\it Physics Department, Ohio State University, Columbus OH 
43210}}
\author{Mark B.\ Wise\footnote{wise@theory.caltech.edu} }
\address{\tighten
{\it California Institute of Technology, Pasadena, CA, USA 91125 }}

\maketitle

{\tighten
\begin{abstract}
Higher order terms in the effective action of noncommutative gauge theories
exhibit generalizations of the $\star$-product ({\it e.g.}  $\star^{\prime}$
and $\star_3$). These terms do not manifestly respect the noncommutative gauge
invariance of the tree level action.  In $U(1)$ gauge theories, we note that
these generalized $\star$-products occur in the expansion of some quantities
that are invariant under noncommutative gauge transformations, but contain an
infinite number of powers of the noncommutative gauge field. One example is an
open Wilson line.  Another is the expression for a commutative field strength
tensor $F_{ab}$ in terms of the noncommutative gauge field $\widehat A_{a}$. 
Seiberg and Witten derived differential equations that relate commutative and
noncommutative gauge transformations, gauge fields and field strengths. In the
$U(1)$ case we solve these equations neglecting terms of fourth order in
$\widehat A$ but keeping all orders in the noncommutative parameter 
$\theta^{kl}$.
\end{abstract}
}
\vspace{0.7in}
\pagebreak

Noncommutative field theories are quantum field theories which live on a
space-time in which the coordinates do not commute:
\begin{eqnarray}\label{com}
[ x^i, x^j ] = i \, \theta^{ij} \,.
\end{eqnarray}
Noncommutative gauge theories emerge naturally when considering the low energy
limit of open strings  in the presence of a background B field. 
\cite{CDS,DH,SW,CK,CH,VS}  The low energy effective action is obtained by
considering the tree level scattering of massless open string states. The zero
slope limit is taken in such a way that the open string metric and
noncommutativity parameter $\theta^{i j}$ remain finite.\cite{SW} In this limit
the effective action for the noncommutative gauge theory can be obtained from
that of the ordinary gauge theory simply by replacing ordinary products of
fields with Moyal or $\star$-products. The $\star$-product is defined by:
\begin{eqnarray}\label{star}
f(x) \star g(x) = \exp \left[\frac{i}{2} \, \theta^{i j} {\partial \over \partial x^i} 
{\partial \over \partial y^j} \right] f(x) g(y) \bigg \bracevert_{x=y} \, .
\end{eqnarray}

For instance, the noncommutative $U(1)$ gauge theory action is
\begin{eqnarray}\label{action}
S = -{1 \over 4 g^2} \int d^4x \, \widehat{F}_{a b} \star \widehat{F}^{a b} \, ,
\end{eqnarray}
where the noncommutative gauge field strength is 
\begin{eqnarray}\label{fieldstrength}
\widehat{F}_{a b} = \partial_a \widehat{A}_b- \partial_b \widehat{A}_a 
- i [ \widehat{A}_a, \star \widehat{A}_b] \, .
\end{eqnarray}
We have introduced the notation $[A, \star B] \equiv A\star B- B \star A$.
In Eq.~(\ref{fieldstrength}), $[A_a, \star A_b]$ does not vanish because the 
$\star$-product is noncommutative. The action in Eq.~(\ref{action}) is 
invariant under the infinitesmal noncommutative gauge transformation
\begin{eqnarray}\label{tran}
\delta_{\widehat{\lambda}} \widehat{A}_a = \partial_a \widehat{\lambda} + 
i [\widehat{\lambda}, \star \widehat{A}_a] \, , \\
\delta_{\widehat{\lambda}} \widehat{F}_{a b} = 
i  [\widehat{\lambda}, \star \widehat{F}_{a b}] \, . \,\,\,\,\,\,\, \nonumber
\end{eqnarray}

Noncommutative field theories exhibit peculiar phenomena unlike that of local
quantum field theories. The action in Eq.~(\ref{action}) is actually thought to
be renormalizable \cite{TF,VG,CDP,MSR,SJ,ABK,GKW,CHMS,EH,CR,GGRS,GS,AA} even
though it contains an infinite number of higher derivative operators.  Another
peculiar phenomenon is UV-IR mixing.  \cite{SRS,RS,MH,FGKPPG,MST} The
commutation relation in Eq.~(\ref{com}) gives rise to an uncertainty relation
which forces objects which are localized over a short distance in one
space direction to be spread out over a long distance in an orthogonal
direction. Thus UV and IR modes of the theory are linked and the usual
decoupling of the ultraviolet from the infrared does not occur in these 
theories.

The appearance of Moyal products in the action of noncommutative gauge theories
is quite natural in light of the commutation relation in Eq.~(\ref{com}). Since
Eq.~(\ref{com}) is essentially identical to the commutation relation of
annihilation and creation operators, there is an isomorphism between functions
$f$ of ordinary coordinates and operators ${\cal O}_f$ in the Hilbert space of
annihilation and creation operators. To define the map from ordinary functions
to operators in Hilbert space, a prescription for specifying the ordering of
annihilation and creation operators in ${\cal O}_f$ is needed. If Weyl ordering
is used to define this map, then it is easy to show that the Moyal product of
functions is isomorphic to ordinary operator multiplication 
(see e.g. \cite{GMS}).

Higher order terms in the effective action for massless open string fields
\cite{LM,Z}, certain global anomalies in $U(1)$ noncommutative gauge
theories coupled to matter \cite{AS} and the coupling of open strings to
closed strings in the presence of a background B field \cite{Garousi,HKLL}
exhibit a more complicated mathematical structure than what is seen at tree
level. Instead of the Moyal products the higher order contributions to the
effective action contain generalized $\star$-products such as
\begin{eqnarray}\label{starprime}
f(x) \star^\prime g(x) = 
{\sin(\frac{\partial_1 \wedge \partial_2}{2})\over 
\frac{\partial_1 \wedge \partial_2}{2}}f(x_1)g(x_2)\bigg\bracevert_{x_i=x}
\end{eqnarray}
and 
\begin{eqnarray}\label{starthree}
[f(x) g(x) h(x)]\star_3 = 
\left[{ \sin(\frac{\partial_2 \wedge \partial_3}{2}) 
\sin(\frac{\partial_1 \wedge (\partial_2 +\partial_3)}{2})
\over \frac{(\partial_1+\partial_2) \wedge \partial_3}{2}
\frac{\partial_1 \wedge (\partial_2+\partial_3)}{2}} 
+ (1 \leftrightarrow 2) \right]
f(x_1)g(x_2)h(x_3)\bigg\bracevert_{x_i=x} \, ,
\end{eqnarray}
where 
\begin{eqnarray}
\partial_1 \wedge \partial_2 = \theta^{i j} {\partial \over \partial x^i_1}
 {\partial \over \partial x^j_2} \, .
\end{eqnarray}
Note that $\star^\prime$ is symmetric in $f$ and $g$ and $\star_3$ is invariant under all permutations of $f,g$ and $h$, though this may not be obvious from 
Eq.~(\ref{starthree}). 

The appearance of these generalized $\star$-products in the effective action is
somewhat confusing. First of all, the effective action does not manifestly
respect the noncommutative $U(1)$ gauge symmetry of Eq.~(\ref{tran}).  It is
hard to believe that radiative corrections do not respect the gauge symmetry of
the tree level action. Second, while the Moyal product can be simply understood
as a consequence of the commutation relation in Eq.~(\ref{com}), the nature of
these generalized $\star$-products remains unclear.  

In this paper, we point out that the generalized $\star$-products also appear
in the expansion of gauge invariant operators constructed in \cite{GHI,DR}. 
These operators are local operators attached to open Wilson lines.  Expanding
the Wilson line to $O(\widehat{A}^2)$ in the gauge field we find the
generalized $\star$-products in Eqs. (\ref{starprime}) and (\ref{starthree}). 
This gives some support to the conjecture of \cite{LM,Z} that there may be 
a way of rewriting the effective action so that the noncommutative
gauge symmetry is manifest. We also point out that a specific gauge invariant
operator is closely related to the solutions of the Seiberg-Witten (SW)
differential equation for the $U(1)$ gauge theory. The solution of SW equation
is a map between ordinary gauge fields and noncommutative gauge fields which
preserves the gauge equivalence classes of the respective theories.  We present
a solution of the SW equation which is correct to all orders in $\theta^{i j}$
and to $O(\widehat{A}^3)$ in the gauge field. This solution also exhibits the
generalized $\star$-products.  

We begin with a brief review of the gauge invariant operators introduced in
\cite{GHI,DR} (see also \cite{IIKK}). Noncommutative $U(1)$ gauge theory with
matter only in the adjoint representation has no local gauge invariant
operators.  There are nonlocal gauge invariant operators which are the Fourier
transform of local operators attached to an open Wilson line. 

The Wilson line is defined by 
\begin{eqnarray}
W(x,C) = P_\star \exp \left(-i \int_0^1 d\sigma {d \zeta^i \over d \sigma}
\widehat{A}_i(x+\zeta(\sigma))\right) ,
\end{eqnarray}
where $C$ denotes path parametrized by $\zeta^i(\sigma)$, such that 
$\zeta^i(0) = 0$ and $\zeta^i(1) = l^i$. $P_\star$ denotes the
usual path ordering with ordinary products of fields replaced by 
$\star$-products. The Wilson line transforms under noncommutative $U(1)$ gauge 
transformations as
\begin{eqnarray}
W(x,C) \rightarrow U(x) \star W(x,C) \star U(x+l)^\dagger \, .
\end{eqnarray}
Now consider a local operator constructed from the noncommutative
gauge field which transforms as
\begin{eqnarray}
 \widehat {\cal O} \rightarrow U(x) \star \widehat {\cal O} \star
U(x)^\dagger \, .
\end{eqnarray}
An example of such an operator is $\widehat {\cal O}=\hat F_{ab}$. The 
operator $\widetilde{{\cal O}}$ with Fourier transform
\begin{eqnarray}
\widetilde{{\cal O}}^{(FT)} = \int d^4x \, \widehat {\cal O} \star W(x,C) 
\star e^{i k x},
\end{eqnarray}
is invariant under noncommutative gauge transformations \cite{GHI,DR}
provided we choose \newline $l^i = \theta^{i j}k_j$. 

Expanding the gauge invariant operator $\widetilde{\cal{O}}$ to 
$O(\widehat{A}^2)$, we obtain
\begin{eqnarray}\label{wl}
\widetilde{{\cal O}}^{(FT)} &=&\int d^4 x \, \widehat {\cal O} \star W(x,C) 
\star e^{i k x} \\
&=& \int d^4x \, \widehat {\cal O} \star
\left[1-i\int_0^1 d\sigma \frac{d\zeta^i}{d\sigma} 
\widehat{A}_i(x+\zeta(\sigma)) \right. \nonumber \\
&& \left. - \int_0^1 d \sigma_1 \int_{\sigma_1}^1 d \sigma_2 
\frac{d\zeta^j}{d\sigma_1}\frac{d\zeta^k}{d\sigma_2} 
\widehat{A}_j(x+\zeta(\sigma_1)) \star \widehat{A}_i(x+\zeta(\sigma_2)) 
+...\right]\star e^{i k x} \, . \nonumber
\end{eqnarray}
The result of doing these integrals depends on the path chosen. We will 
choose the path corresponding to a straight Wilson line:
$\zeta^i(\sigma) = \theta^{i j} k_j \sigma$. We then use
\begin{eqnarray}\label{trick}
\widehat{A}_i(x+\theta k \sigma) = e^{-\sigma k \wedge \partial} 
\widehat{A}_i(x).
\end{eqnarray}
The integrals are over $\sigma_i$ are simple and it is not hard to show
that 
\begin{eqnarray}\label{wl2}
\widetilde{{\cal O}}^{(FT)} = \int d^4x \left[\widehat {\cal O}
+ \theta^{i j}\partial_j(\widehat {\cal O} \star^\prime \widehat{A}_i)
+\frac{1}{2} \theta^{i j}\theta^{k l}\partial_j \partial_l
[\widehat {\cal O}\, \widehat{A}_i \,\widehat{A}_k]\star_3 +...\right]
e^{i k x} \, .
\end{eqnarray}
It is straightforward to check that the series inside the brackets
of Eq.~(\ref{wl2}) is gauge invariant up to terms of $O(\widehat{A}^2)$. Two 
identities which are useful for this are
\begin{eqnarray}\label{trick2}
\theta^{i j} \partial_i f \star^\prime \partial_j g = -i [f, \star g] \,,
\,\,\,\,\,\,\,\,\,\,\,\,\,\,\,\,\,\,\,\,\,\, \\
f\star^\prime [g, \star h] + g\star^\prime [f, \star h] =
i \theta^{i j} \partial_i [f \, g \, \partial_j h]\star_3 \, .\nonumber
\end{eqnarray}
Even though the definition of $\widetilde O$ was given in terms of ordinary $\star$-products its expansion in powers of the noncommutative
gauge field involves $\star^{\prime}$- and $\star_3$-products.

Next we briefly discuss the Seiberg-Witten map between ordinary gauge fields
and noncommutative gauge fields and show that the generalized $\star$-products also appear in this map.

Noncommutative Yang-Mills is in fact equivalent to ordinary gauge theory
perturbed by an infinite number of higher dimension operators. It is possible
to show that there exists a map between noncommutative gauge fields
and ordinary gauge fields which preserves the gauge equivalence classes of the
respective theories, despite the fact that the theories have two different
gauge groups \cite{SW}. One way to demonstrate this is to show that the two 
space-time theories can be obtained from the same world sheet sigma model 
regulated in two different ways.

The relationship between the commutative gauge field and field strength and the
analogous noncommutative quantities depends on the parameter $\theta^{kl}$. For
$\theta^{kl} =0$, $F_{a b}=\widehat{F}_{a b}$ and $A_a = \widehat{A}_a$.  It is possible derive differential equations for the gauge field, gauge parameter and field strength as functions of $\theta^{kl}$: \cite{SW}
\begin{eqnarray}\label{SeiWit}
\delta \widehat{A}_a = -{1\over 4} \delta \theta^{kl} 
\left[\widehat{A}_k \star (\partial_l \widehat{A}_a + \widehat{F}_{l a})
+ (\partial_l \widehat{A}_a + \widehat{F}_{l a}) \star \widehat{A}_k \right] 
\, , \,\,\,\,\,\,\,\,\,\,\,\,\,\,\,\,\,\,\,\,\,\,\,\,\,\,\,\,\,\, \\
\delta \widehat{\lambda} = {1\over 4} \delta \theta^{kl} 
(\partial_k \widehat{\lambda} \star \widehat{A}_l + \widehat{A}_l \star 
\partial_k \widehat{\lambda} ) \, , \,\,\,\,\,\,\,\,\,\,\,\,\,\,\,\,\,\,\,\,
\,\,\,\,\,\,\,\,\,\,\,\,\,\,\,\,\,\,\,\,\,\,\,\,\,\,\,\,\,\,\,\,\,\,\,\,\,\,
\,\, \nonumber \\
\delta \widehat{F}_{a b} = {1\over 4} \delta \theta^{kl} 
\left[ 2 \widehat{F}_{a k} \star \widehat{F}_{b l} + 2 \widehat{F}_{b l} \star 
\widehat{F}_{a k}- \widehat{A}_k \star (\widehat{D}_l \widehat{F}_{a b} + 
\partial_l \widehat{F}_{a b}) - (\widehat{D}_l \widehat{F}_{a b} + \partial_l 
\widehat{F}_{a b}) \star \widehat{A}_k \right]. \nonumber
\end{eqnarray}
In the remainder of this paper, we will refer to these equations as the 
Seiberg-Witten (SW) equations.

In \cite{SW} the SW equations were solved for the special case of constant
$U(1)$ fields. In \cite{Garousi}, the equations were solved to
$O(\widehat{A}^2)$ by integrating along a special path in the space of matrices
$\theta^{kl}$. Specifically, \cite{Garousi} takes the anticommuting parameter 
to be $\alpha \, \theta^{kl}$, then integrates $\alpha$ from 0 to 1 with the 
boundary condition $\widehat{F}_{a b} (\alpha = 0) = F_{a b}$, $\widehat{A}_{a} (\alpha = 0) = A_{a}$ and $\widehat{\lambda}(\alpha =0) = \lambda$.  
Note that the solution obtained by integrating the SW equation
depends on the path of integration \cite{AI}. However, the path dependence can
be absorbed  entirely into a field redefinition of the noncommutative gauge
field.  For the case of a $U(1)$ gauge theory, the solution for the field 
strength tensor is
\begin{eqnarray}\label{soln2} 
F_{a b} = \widehat{F}_{a b} + \theta^{k l}(\widehat{A}_k \star^\prime 
\partial_l \widehat{F}_{a b} - \widehat{F}_{a k}\star^{\prime}\widehat{F}_{b l})
+ O(\widehat{A}^3)  \, . 
\end{eqnarray} 
The result is exact in $\theta$ and correct to order $O(\widehat{A}^2)$ in the
fields. The left hand side of Eq.~(\ref{soln2}) is invariant under the ordinary
$U(1)$ gauge transformation. Since the SW map preserves gauge equivalence
classes, the right hand side of Eq.~(\ref{soln2}) should be invariant 
under noncommutative $U(1)$ gauge transformations to 
$O(\widehat{A})$. It is straightforward to check that this is the case.

Integrating the SW equations to $O(\widehat{A}^2)$ is straightforward because
at this order one can neglect the $\theta^{kl}$ dependence of $\widehat{A}_a$
and $\widehat{F}_{ab}$ in Eq.~(\ref{SeiWit}). Direct integration of the SW
equations is difficult at higher orders in $\widehat{A}$. Below we will give a
method which allows one to obtain higher order solutions to the SW equations.

This method exploits the similarity of the solution to the SW equation and 
the gauge invariant operator:
\begin{eqnarray}\label{F2}
\widetilde{F}_{ab}^{(FT)} &=& \int d^4 x \widehat{F}_{a b} \star W(x,C) \star
e^{i k x} \\
&=& \int d^4x \left[ \widehat{F}_{a b} + \theta^{k l} \partial_l 
(\widehat{A}_k \star^\prime \widehat{F}_{a b}) + ... \right] e^{i k x}  
\nonumber \\ &=& \int d^4x \left[ \widehat{F}_{a b} + \theta^{k l}
(\widehat{A}_k \star^\prime\partial_l \widehat{F}_{a b}
+ \frac{1}{2} \widehat{F}_{l k} \star^\prime \widehat{F}_{a b} ) +... 
\right] e^{i k x} \, . \nonumber 
\end{eqnarray}
Notice that one of the order $O(\widehat{A}^2)$ terms in the expansion of
$\widetilde{F}_{a b}$ is identical to an $O(\widehat{A}^2)$ term in the
solution to the SW equation for $F_{a b}$ given in
Eq.~(\ref{soln2}). This is because both $\widetilde{F}_{a
b}$ and $F_{a b}$ are gauge invariant, and the term $\theta^{k l} \widehat{A}_k
\star^\prime\partial_l \widehat{F}_{a b}$ is necessary to ensure gauge
invariance to $O(\widehat{A})$.  $\widetilde{F}_{a b}$ and
$F_{a b}$ differ by terms which are by themselves gauge invariant to
$O(\widehat{A})$. A crucial difference between $\widetilde{F}_{a b}$ and
$F_{ab}$ is that $F_{a b}$ is a field strength and therefore obeys the Bianchi
identity while $\widetilde{F}_{a b}$ does not.  

Using what we know about the gauge invariant operator $\widetilde{F}_{a b}$ it
is possible to construct the $O(\widehat{A}^3)$ solution to the SW equations. 
To the $O(\widehat{A}^2)$ solution to the SW equations for the field
strength given in Eq.~(\ref{soln2}) we add
the $O(\widehat{A}^3)$ term in the expansion of $\widetilde{F}_{a b}$
\begin{eqnarray}\label{st1}
\widehat{F}_{a b} +\theta^{i j}\partial_j(\widehat{A}_i \star^\prime 
\widehat{F}_{a b}) + \frac{1}{2} \theta^{i j}\theta^{k l}\partial_i 
\partial_k [\widehat{F}_{a b} \, \widehat{A}_j \, \widehat{A}_l]\star_3 
+\theta^{i j}\left(\frac{1}{2}\widehat{F}_{a b}\star^\prime \widehat{F}_{i j} - 
\widehat{F}_{a i} \star^\prime \widehat{F}_{b j}\right)\, .
\end{eqnarray}
The first three terms are gauge invariant to $O(\widehat{A}^2)$. The remaining
terms under a gauge transformation give rise to an $O(\widehat{\lambda} 
\widehat{F}^2)$ term, which can be cancelled by adding terms 
$O(\widehat{A} \widehat{F}^2)$. Using the identities in Eq.~(\ref{trick2}) it 
is easy to show that the following quantity is gauge invariant to 
$O(\widehat{A}^2)$:
\begin{eqnarray}\label{githree}
\widehat{F}_{a b} &+&
\theta^{i j}\bigg( \partial_j (\widehat{A}_i \star^\prime \widehat{F}_{a b}) + 
\frac{1}{2} \widehat{F}_{a b}\star^\prime \widehat{F}_{i j} 
-\widehat{F}_{a i}\star^\prime \widehat{F}_{b j} \bigg) \\ 
&+& \frac{1}{2}\theta^{i j} \theta^{k l}\bigg(\partial_i \partial_k 
[\widehat{F}_{a b} \, \widehat{A}_l \, \widehat{A}_j]\star_3 
- \partial_k [\widehat{F}_{i j} \, \widehat{F}_{a b} \, \widehat{A}_l]\star_3 
+ 2 \partial_k [\widehat{F}_{a i} \, \widehat{F}_{b j} \, \widehat{A}_l]\star_3 \bigg) \, . \nonumber
\end{eqnarray}
Eq.~(\ref{githree}) can be written as 
\begin{eqnarray}
\theta^{i j}\theta^{k l}\bigg(\frac{1}{2} [\widehat{F}_{a i}\widehat{F}_{b j}
\widehat{F}_{k l}]\star_3- \frac{1}{8}[\widehat{F}_{a b}\widehat{F}_{i j}
\widehat{F}_{k l}]\star_3- \frac{1}{4}[\widehat{F}_{a b}\widehat{F}_{j k}
\widehat{F}_{i l}]\star_3 + [\widehat{F}_{i k}\widehat{F}_{a l}\widehat{F}_{b j}]\star_3\bigg)
+\partial_a A_b - \partial_b A_a ,
\end{eqnarray}
where 
\begin{eqnarray}\label{ncgf}
A_b &=& \widehat{A}_b +\frac{1}{2}\theta^{i j} \widehat{A}_i \star^\prime (\partial_j \widehat{A}_b +\widehat{F}_{j b}) \\ 
&& + \frac{1}{2}\theta^{i j}\theta^{k l} [- \widehat{A}_i \, \partial_k 
\widehat{A}_b \,(\partial_j \widehat{A}_l + \widehat{F}_{j l}) + \partial_k 
\partial_i \widehat{A}_b \, \widehat{A}_j \, \widehat{A}_l + 2 \partial_k 
\widehat{A}_i \, \partial_b \widehat{A}_j \, \widehat{A}_l]\star_3 
+ O(\widehat{A}^4). \nonumber
\end{eqnarray}
Since Eq.~(\ref{githree}) is gauge invariant to $O(\widehat{A}^2)$, $A_b$ 
must transform like an ordinary $U(1)$ gauge field up to terms 
$O(\widehat{A}^3)$. Hence, $A_b$ is a solution to the SW differential equation
for the gauge field. Explicitly $\delta_{\lambda} A_b = \partial_b \lambda$ 
where, 
\begin{eqnarray}\label{gpr}
\lambda 
= \widehat{\lambda} +
\frac{1}{2}\theta^{i j}\widehat{A}_i \star^\prime \partial_j \widehat{\lambda}
+\frac{1}{2}\theta^{i j}\theta^{k l} [\partial_k \partial_i \widehat{\lambda}
\, \widehat{A}_j \, \widehat{A}_l +\partial_k \widehat{\lambda} \, 
\widehat{A}_i \, \partial_l \widehat{A}_j]\star_3 +O(\widehat{A}^3).
\end{eqnarray}
$\lambda$ in Eq.~(\ref{gpr}) is the solution to the SW differential equation
for the gauge parameter. The field strength constructed from the gauge field in
Eq.~(\ref{ncgf}),  $F_{a b} = \partial_a A_b -\partial_b A_a$, is gauge
invariant to $O(\widehat{A}^2)$ and obeys the Bianchi identity. Using our
previous results it is easy to show that
\begin{eqnarray}\label{ncfs}
F_{ab}&=&\widehat{F}_{a b} +
\theta^{i j}\bigg( \partial_j (\widehat{A}_i \star^\prime \widehat{F}_{a b}) + 
\frac{1}{2} \widehat{F}_{a b}\star^\prime \widehat{F}_{i j} 
-\widehat{F}_{a i}\star^\prime \widehat{F}_{b j} \bigg) \\ \nonumber
&&+ \frac{1}{2}\theta^{i j} \theta^{k l}\bigg(\partial_i \partial_k 
[\widehat{F}_{a b} \, \widehat{A}_i \, \widehat{A}_j]\star_3 
- \partial_k [\widehat{F}_{i j} \, \widehat{F}_{a b} \, \widehat{A}_l]\star_3 
+ 2 \partial_k [\widehat{F}_{a i} \, \widehat{F}_{b j} \, \widehat{A}_l]\star_3 \bigg)\\
&&-\theta^{i j}\theta^{k l}\bigg(\frac{1}{2} [\widehat{F}_{a i}\widehat{F}_{b j}
\widehat{F}_{k l}]\star_3- \frac{1}{8}[\widehat{F}_{a b}\widehat{F}_{i j}
\widehat{F}_{k l}]\star_3- \frac{1}{4}[\widehat{F}_{a b}\widehat{F}_{j k}
\widehat{F}_{i l}]\star_3 + [\widehat{F}_{i k}\widehat{F}_{a l}\widehat{F}_{b j}]\star_3\bigg) + O(\widehat{A}^4) \, .\nonumber
\end{eqnarray}
$F_{ab}$ in Eq.~(\ref{ncfs}) is the solution of the SW differential
equation for the field strength.

In \cite{KO} a path integral representation of the map between commutative and
noncommutative gauge fields is derived. Explicit constructions of the
Seiberg-Witten map have also been obtained in \cite{JS,JSW} and extended to the
non-abelian case in \cite{JSSW}. In \cite{KO}, a solution for $\widehat{A}(A)$
to $O(A^3)$ is obtained. The $O(A^2)$ part of the solution in \cite{KO} is
exact in $\theta$ and exhibits the $\star^\prime$-product. However, the
$O(A^3)$ part of the solution in \cite{KO} is not exact in $\theta$ so does not
have the $\star_3$-product.

The SW equations only involve ordinary $\star$-products. However we have seen
that the solution to these equations when expanded in powers of the
noncommutative gauge field involve $\star^{\prime}$- and $\star_3$-products. 
These generalized $\star$-products also appear in the expansion of the gauge
invariant Wilson line, as well higher orders in the effective action of
noncommutative gauge theories and in the coupling of massless closed string
states to noncommutative gauge fields.  Though existing calculations
\cite{LM,Z} of higher order terms in the effective action do not respect the
noncommutative gauge symmetry of the tree level action, the existence of
similar structures in the expansion of gauge invariant nonlocal quantities
suggests that it should be possible to write down gauge invariant, albeit
non-local, expressions for the effective action. \cite{LM,Z}

We thank J. Gomis and E. Witten for useful discussions.  T.M. was supported by
the National Science Foundation under Grant No. PHY-9800964.  M.B.W. was
supported in part by the Department of Energy under grant DE-FG03-ER-40701.

\end{document}